\documentclass[aps,pla,twocolumn,10pt,showpacs,superscriptaddress,footnoteinbib]{revtex4}
\usepackage{amsmath}
\usepackage{latexsym}
\usepackage{amssymb}
\usepackage{graphics,epstopdf}
\begin{document}

\title{A Testable Prediction of the No-signalling Condition using a variant of the EPR-Bohm example}
\author{Dipankar Home}
\email{dhome@bosemain.boseinst.ac.in}
\affiliation{CAPSS, Department of Physics, Bose Institute, Sector-V, Salt Lake, Kolkata 700 091, India}
\author{Ashutosh Rai}
\email{arai@bose.res.in}
\affiliation{S. N. Bose National Centre for Basic Sciences, Block JD, Sector-III, Salt Lake, Kolkata 700 098, India}
\author{A. S. Majumdar}
\email{archan@bose.res.in}
\affiliation{S. N. Bose National Centre for Basic Sciences, Block JD, Sector-III, Salt Lake, Kolkata 700 098, India; Tel: +91-33-2335-5706; 
Fax: +91-33-2335-3477}
\date{\today}
\maketitle

\vskip 0.5cm

\begin{center}
\large{\bf Abstract}
\end{center}
Predictive power of the no-signalling condition (NSC) is demonstrated in a \textit{testable} situation involving a non-ideal Stern-Gerlach (SG)
device in one of the two wings of the EPR-Bohm entangled pairs. In this wing, for two types of measurement in the other wing, we consider the spin state of a selected set of particles
that are confined to a particular half of the
plane  while emerging from the SG magnetic field region. Due to 
non-idealness of the SG setup, this spin state will have superposing components involving a relative phase for which a testable \textit{quantitative constraint} is obtained by  invoking NSC, thereby providing a means for precision testing of this fundamentally significant principle.

\section{Introduction} 
A key condition underpinning the `peaceful coexistence' \cite{shimony} between quantum mechanics and special relativity is the no-signalling condition (NSC) which prohibits the use of quntum nonlocality for sending information in a way that can lead to causality paradoxes. While the compatibility of NSC with quantum mechanics has been extensively analysed with an increasing generality \cite{bohm}, an interesting line of study was initiated by Gisin showing the use of NSC as a tool to either find the limits of quantum mechanics, like constraining any conceivable nonlinear modification of the Schr\"{o}dinger equation \cite{gisin}, or to obtain specific bounds on quantum operations, like deriving bound on the fidelity of quantum cloning machines \cite{gisin1}. Subsequently, NSC has been applied, for example, to obtain a tight bound on  the optimal unambiguous discrimination between two nonorthogonal states \cite{barnett}. A number of other studies \cite{feng} too have highlighted the role of NSC in limiting various quantum operations, while some features of the quantum formalism have also been derived from NSC \cite{svetlichny}. Interestingly, NSC has been invoked in the context of quantum cryptography as well; viz.  to formulate quantum key distribution protocols which are secure against attack by any  eavesdropper who is limited only by the impossibility of superluminal signalling \cite{barrett} - this line of study is motivated by the notion \cite{barrett1}: ``... quantum theory could fail without violating standard relativistic causality, and vice versa''.

Complementing the above mentioned studies, our present work aims at deriving from NSC, a \textit{testable quantitative relation} whose empirical scrutiny would enable a dedicated precision testing of NSC in the same spirit as Born's rule has recently been tested to provide bounds on its accuracy \cite{urbashi}. Such precision testing using a quantitative relation is useful in having the potential of being able to detect very small deviations that may be missed otherwise, and can provide an empirical upper bound on possible violation of NSC. To the best of our knowledge, a dedicated precision testing of NSC remains unexplored. To this end, the example analysed here provides a constraint relation concerning the spin state of a selected set of spin-1/2 particles that are confined to a particular half of the plane while emerging from a \textit{non-ideal} Stern-Gerlach(SG) setup. Here the particles in the other half of the plane are taken to be blocked/detected. The spin state, thus, filtered out comprises of two superposing spin components with a relative phase, the respective probability amplitudes being calculable from the solutions of the Schr\"{o}dinger equation for the non-ideal SG setup. In order to evaluate the relative phase, a fully unitary treatment (albeit non-trivial) is required that would be based on appropriately modeling the post-selection process used for filtering out the spin state. The central point made in this paper is that irrespective of the specifics of such a treatment, the value of the relative phase occurring in the post-selected spin state has to satisfy a testable constraint that is derivable from NSC. To show this, we proceed as follows.

\section{Formulation of the example} 
For the EPR-Bohm (EPRB) entangled pairs of spin-1/2 particles in spin singlets, the corresponding wave function is given by

\begin{equation}
\label{t1}
|\Psi\rangle=(1/\sqrt{2})|\psi_{0} \rangle_{1}|\psi_{0} \rangle_{2} (|\uparrow\rangle_{1} |\downarrow \rangle_{2}- |\downarrow\rangle_{1} |\uparrow\rangle_{2})
\end{equation}

where the spatial parts $|\psi_{0}\rangle_{1}$ and $|\psi_{0}\rangle_{2}$ (assumed to be, say, represented by Gaussian wave packets) pertain to the particles 1 and 2 respectively, and the spin part represents the singlet state. Next, let a \textit{non-ideal} SG setup be placed by Bob in one of the two wings of the EPRB pairs, say, in the wing 2 for the particles moving along the +y-axis (Fig. 1). After passing through an inhomogeneous magnetic field in the SG setup oriented along, say, the +z-axis, the time-evolved total wave function ($\Psi(\textbf{x}, t)$) of any particle, in general, would involve the spatial wave functions $\psi_{+}(\textbf{x}, t)$ and  $\psi_{-}(\textbf{x}, t)$ that are coupled with the spin-up $(|\uparrow\rangle_{z})$ and spin-down $(|\downarrow\rangle_{z})$ states respectively, given by

\begin{equation}
\label{t1a}
\Psi(\textbf{x}, t) = \psi_{+}(\textbf{x}, t)|\uparrow\rangle_{z} + \psi_{-}(\textbf{x}, t)|\downarrow\rangle_{z}
\end{equation}

Here $|\psi_{+}(\textbf{x}, t)|^{2}$ $(|\psi_{-}(\textbf{x}, t)|^{2})$ determines the probability of finding particles with the spin $|\uparrow\rangle_{z} (|\downarrow\rangle_{z})$ in the upper and lower halves of the y-z plane. In the \textit{ideal} case, $\langle\psi_{+}(\textbf{x}, t)|\psi_{-}(\textbf{x}, t)\rangle=0$ with the probability of finding particles corresponding to the spin $|\uparrow\rangle_{z} (|\downarrow\rangle_{z})$ in the lower (upper) y-z plane being negligibly small. The explicit forms of $\psi_{+}(\textbf{x}, t)$ and $\psi_{-}(\textbf{x}, t)$ in the general case of a \textit{non-ideal} SG setup are available in the relevant literature (see, for example, \cite{home, pan} and the references cited therein), where $\langle\psi_{+}(\textbf{x}, t)|\psi_{-}(\textbf{x}, t)\rangle\neq0$. Note that idealness (non-idealness) of a SG setup depends on appropriately choosing the strength of the magnetic field within the SG setup, commensurate with the energy and width of the incoming wave packet. 

Now, suppose two types of measurements (the types denoted by A and B respectively) are performed by, say, Alice in the wing 1 of the EPRB pairs of singlet states given by Eq.(\ref{t1}). The type A corresponds to a set of measurements of, say, the x-component of spin, while B corresponds to a set of measurements of the spin component along the z-axis which is the same as the direction of the inhomogeneous magnetic field in the non-ideal SG setup in Bob's wing. Therefore, in the cases A and B respectively, effectively mixtures of $+x$ and $-x$ spin components (with equal weighting)and that of $+z$ and $-z$ spin components (with equal weighting) are produced in Bob's wing. This can be seen from Eq. (\ref{t1}) by either invoking the collapse postulate or by taking into account the feature that the measuring apparatus states in Alice's wing are mutually orthogonal, corresponding to the distinct outcomes of the measurement of the spin. Thus, subsequently in Bob's wing, say, in the 
 case A, for the purpose of our argument, we can consider the spin states $|\rightarrow\rangle_{x}$ and $|\leftarrow\rangle_{x}$ as pure state constituents of a mixed state. Similar is what happens in the case B where the resulting mixed state comprises of $|\uparrow\rangle_{z}$ and $|\downarrow\rangle_{z}$ states. The particles in Bob's wing are then passed through a non-ideal SG setup, followed by post-selection confined to upper half of the y-z plane. Such post-selected particles are subjected to the measurement of an arbitrary component of spin (say, $\sigma_{\theta}$) using  an ideal SG setup with its inhomogeneous magnetic field oriented along a direction making an angle $\theta$ with the $+z$-axis in the x-z plane.

\begin{figure}
\resizebox{8cm}{5cm}{\includegraphics{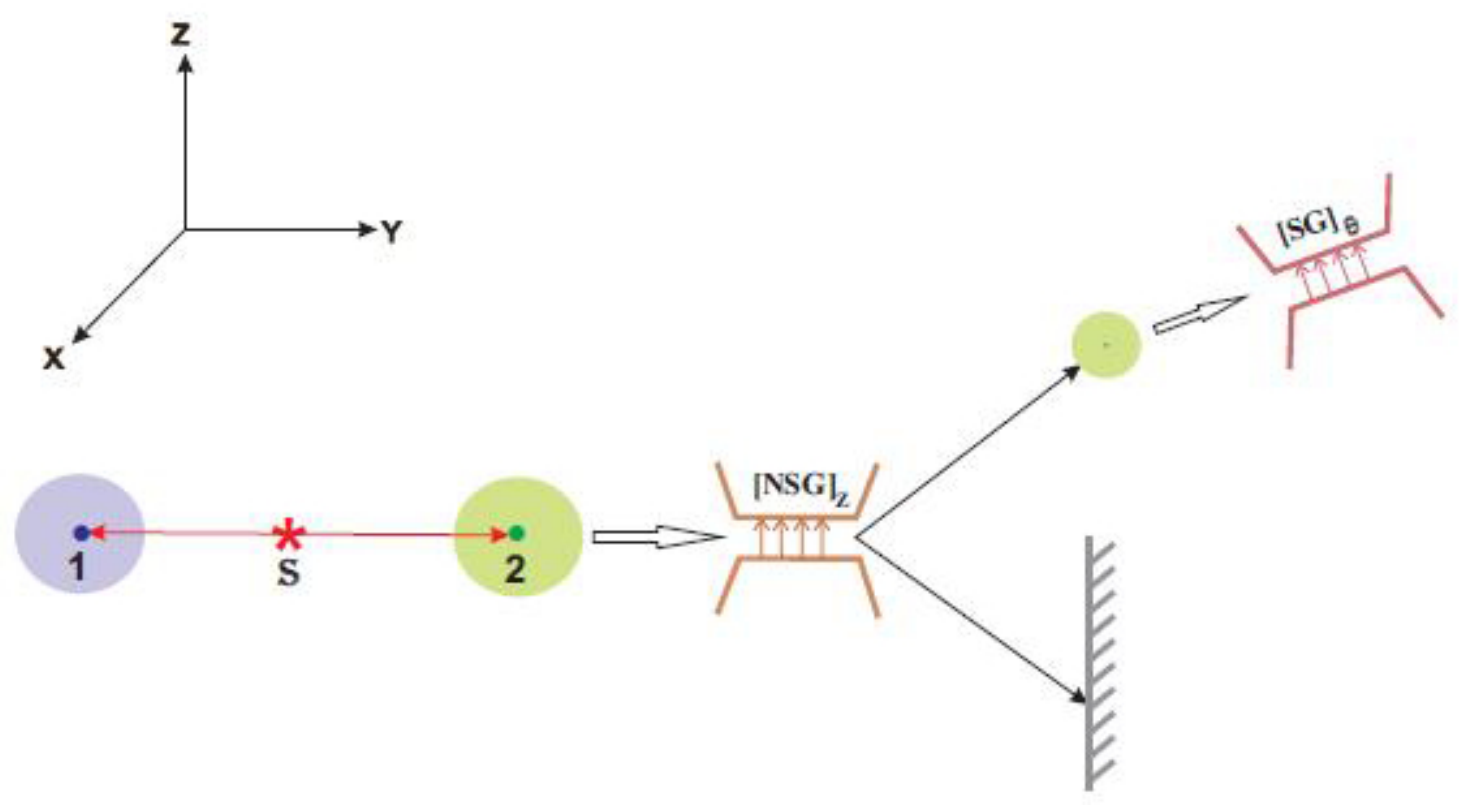}}
\caption{Spin-1/2 particles 1 and 2 are members of the EPR entangled pairs emitted from a source \textbf{S} moving along y-axis in the opposite directions. Particles 2 pass through $[NSG]_{z}$, a nonideal Stern-Gerlach device with its inhomogeneous magnetic field oriented along z-axis. After emerging from the $[NSG]_{z}$ setup, particles confined to the lower half of the y-z plane are absorbed/detected, while particles in the upper half of the y-z plane are subjected to the measurement of an arbitrary spin component, say, $\sigma_{\theta}$ by using an ideal Stern-Gerlach setup  $[SG]_{\theta}$  where the inhomogeneous magnetic field is along a direction in the x-z plane making an angle $\theta$ with the z-axis.}
\end{figure}

Our treatment hereafter will be focused on the probability of obtaining a particular outcome (say, +1) of the measurement of $\sigma_{\theta}$ on the post-selected particles in the upper half of the y-z plane in Bob's wing, calculating it in the two cases A and B corresponding to different types of measurement performed in Alice's wing. Such an observable probability needs to be the \textit{same} in both these cases in order to ensure no-signalling between the two wings of the EPRB pairs. This requirement leads to a constraint on the relative phase occurring in the spin state of the post-selected particles on which the final spin measurement is considered.

\section{Derivation of a testable consequence of the no-signalling condition} 

We begin by noting that post-selecting particles emerging in the upper half of the SG setup in question is ensured by absorbing/detecting particles in the lower half. Thus, this can be viewed, in principle, as a kind of approximate measurement of position of the particles lying within the lower half, for which the spatial states of the lower half particles are \textit{coupled} with the states of the abosorber/detector. Hence, this post-selection process \textit{decoheres} the entanglement (Eq.(\ref{t1a})) between the spatial and spin states of the particles, and results in a product wave function of the spatial and spin parts for all the upper (lower) half particles that emerge from the non-ideal SG setup with its inhomogeneous magnetic field oriented along the z-axis. The spin state of any such particle will be found to be either $|\uparrow\rangle_{z}$ or $|\downarrow\rangle_{z}$, but since it is not known a priori which particle is to be found in which of these states, the spin part of the state is to be regarded as a superposition of $|\uparrow\rangle_{z}$ and $|\downarrow\rangle_{z}$ spin states. The post-selection procedure that is considered may, therefore, be viewed as a state preparation for futher measurements.

In what follows, the cases A and B are analysed separately.

\textbf{Case A:} In this case, as mentioned earlier, due to measurements in Alice's wing, effectively a mixed state comprising of $|\rightarrow\rangle_{x}$ and $|\leftarrow\rangle_{x}$ spin components (with equal weighting) results in Bob's wing. Let us first focus on particles with the $|\rightarrow\rangle_{x}$ spin component in Bob's wing. Given that such particles are passed through a \textit{non-ideal} SG setup involving an inhomogeneous magnetic field along the z-axis, the spin component of any emerging particle  post-selected in the upper half of the y-z plane will be found to be either $|\uparrow\rangle_{z}$ or $|\downarrow\rangle_{z}$ with the respective probability determined by the overlap of $|\psi_{+}(\textbf{x}, t)|^{2}$ or $|\psi_{-}(\textbf{x}, t)|^{2}$ in the spatial region of the upper y-z plane, while, as mentioned earlier, one does not know a priori which  particle will be found in which spin state. Since our subsequent argument is concerned solely with calculating the observational results pertaining to the spin state of the post-selected particles, we can henceforth ignore the spatial  state.
The normalized spin state of any particle belonging to a post-selected ensemble in the upper half will be a superposition of $|\uparrow\rangle_{z}$ and $|\downarrow\rangle_{z}$ spin states with a relative phase denoted by $\phi_{+x}$ (the subscript +x is used for the $|\rightarrow\rangle_{x}$ spins passed through the non-ideal SG setup), given by

\begin{equation}
\label{t2}
|\chi\rangle=\sqrt{1-E_{s}}|\uparrow\rangle_{z}+exp(i\phi_{+x})\sqrt{E_{s}}|\downarrow\rangle_{z}
\end{equation}

\noindent where $E_{s}$ denotes the
time-saturated value of the quantity E(t) given by

\begin{subequations}
\begin{equation}
\label{t3a}
E(t)=\int_{x\rightarrow-\infty}^{+\infty}\int_{y\rightarrow-\infty}^{+\infty}\int_{z=0}^{+\infty}|\psi_{-}(\textbf{x}, t)|^{2}dx dy dz
\end{equation}

\begin{equation}
\label{t3b}
= \int_{x\rightarrow-\infty}^{+\infty}\int_{y\rightarrow-\infty}^{+\infty}\int_{z\rightarrow-\infty}^{+0}|\psi_{+}(\textbf{x}, t)|^{2}dx dy dz
\end{equation}
\end{subequations}

A measure of the non-idealness of the SG setup is provided by the quantity E(t) which determines the probability of finding $|\downarrow\rangle_{z}$ ($|\uparrow\rangle_{z}$) particles in the upper (lower) y-z plane at time t. The parameter E(t) varies with time as the wave packets $|\psi_{+}(\textbf{x}, t)|^{2}$ and $|\psi_{-}(\textbf{x}, t)|^{2}$ freely propagate in opposite directions after emerging from the SG setup. E(t) finally attains a time-independent saturated value denoted by $E_{s}$, with the saturation time depending upon choices of the relevant parameters (see, for example, Home et al. \cite{home}). For an \textit{ideal} SG setup, $E_{s}=0$. 

Here it needs to be pointed out that , although in our argument we are considering, in principle, the post-selection to be done on the whole upper half of the y-z plane, in practice, it would suffice if a representative set of a suitably large number of particles are post-selected that are confined to a finite region of the upper y-z plane such that there is sufficient non-vanishing probability for finding either $|\uparrow\rangle_{z}$ or $|\downarrow\rangle_{z}$ spin state in the half under consideration. Note that due to Gaussian nature of the wave packets $|\psi_{+}|^2$ and $|\psi_{-}|^2$, the probability for finding particles with either $|\uparrow\rangle_{z}$ or $|\downarrow\rangle_{z}$ spin state in the region corresponding to large
z would be negligibly small. Of course, if one is required to take into account the finiteness of the region chosen for post-selection, the range of integration over the z-coordinate in Eqs. (\ref{t3a}), (\ref{t3b}) will vary depending upon the actual region of the upper y-z plane chosen for post-selection, thereby affecting the value of the parameter $E_{s}$ in Eq. (\ref{t2}). But, it is important to stress that, whatever be the value of $E_{s}$, our subsequent argument goes through and the constraint relation concerning the relative phase we now proceed to derive would remain unaffected. 

Now, given an input spin state $|\rightarrow\rangle_{x}=(1/\sqrt{2})(|\uparrow\rangle_{z}+|\downarrow\rangle_{z})$ passing through a non-ideal SG setup, using Eq. (\ref{t1}), the probability for finding particles in the upper half of the y-z plane with the spin component $|\uparrow\rangle_{z}$ or $|\downarrow\rangle_{z}$ is given by $(1/2)(1-E_{s})$ or $(1/2)E_{s}$ respectively. Then, the total probability of such particles being post-selected in the upper half of the y-z plane is given by $p_{upper} = (1/2)(1-E_{s})+(1/2)E_{s}=1/2$. For these post-selected particles with the normalized spin state $|\chi\rangle$ given by Eq. (\ref{t2}), using the expressions for the spin states $|\uparrow\rangle_{z}$ and $|\downarrow\rangle_{z}$ in terms of the eigenstates of $\sigma_{\theta}$ with eigenvalues +1 and -1 respectively, the probability of obtaining a particular outcome, say +1, for the measurement of an arbitrary spin component $\sigma_{\theta}$ is given by

\begin{equation}
\label{t4}
p_{A}^{+} = (1/2)\left[1 + (1-2E_{s})cos\theta + 2\sqrt{E_{s}(1-E_{s})} sin\theta cos\phi_{+x}\right]
\end{equation}

Next, note that due to measurements of the x-component of spin in Alice's wing, particles with the $|\rightarrow\rangle_{x}$ spin component are produced in Bob's wing with the probability (1/2). Hence, the total probability $P_{A}^{+}$ that such particles after passing through a non-ideal SG setup get post-selected in the upper half of the y-z plane and yield the outcome +1 for the measurement of $\sigma_{\theta}$ is given by $P_{A}^{+}=(1/2)p_{upper}  p_{A}^{+}$ where $p_{A}^{+}$ is given by Eq. (\ref{t4}) and $p_{upper}=1/2$ as explained earlier, whence

\begin{equation}
\label{t5}
P_{A}^{+} = (1/8)\left[1+(1-2E_{s})cos\theta + 2\sqrt{E_{s}(1-E_{s})}sin\theta cos\phi_{+x}\right]
\end{equation}

The other situation in the case A that occurs with the probability (1/2) corresponds to particles with the $|\leftarrow\rangle_{x}$ spin component produced in Bob's wing due to measurements of the x-component of spin in Alice's wing. The probability $P_{A}^{-}$ of such particles to be post-selected in the upper half of the y-z plane of the \textit{non-ideal} SG setup and yield the outcome +1 for the measurement of $\sigma_{\theta}$ is given by (obtained in a way similar to the derivation of the expression (\ref{t5}) for the quantity $P_{A}^{+}$)

\begin{equation}
\label{t6}
P_{A}^{-} = (1/8)\left[1+(1-2E_{s})cos\theta + 2\sqrt{E_{s}(1-E_{s})}sin\theta cos\phi_{-x}\right]
\end{equation}

\noindent where $\phi_{-x}$ is the relative phase occurring in the spin state (of the form Eq. (\ref{t2})) of such post-selected particles in the upper half of the y-z plane for the $|\leftarrow\rangle_{x}$ spins passed through the SG setup. Therefore, in the case A, using Eqs. (\ref{t5}) and (\ref{t6}), the total probability $P_{A}^{x}$ of obtaining the outcome +1 for the measurement of $\sigma_{\theta}$ on particles that pass through the non-ideal SG setup and get selected in the upper half of the y-z plane is given by 

\begin{eqnarray}\nonumber
\label{t7}
P_{A}^{x}=P_{A}^{+}+P_{A}^{-}=(1/4)\left[1+(1-2E_{s})cos\theta+\sqrt{E_{s}(1-E_{s})}\right.\\
\left. sin\theta (cos\phi_{+x}+ cos\phi_{-x})\right]
\end{eqnarray}

\noindent where the superscript $x$ is used to denote that the quantity $P_{A}^{x}$ is measured in Bob's wing corresponding to the total set of measurements of the x-component of spin in Alice's wing.

\textbf{Case B:} In this case, we consider a set of measurements of the z-component of spin in Alice's wing which is along the \textit{same} direction as that of the inhomogeneous magnetic field in the non-ideal SG setup in Bob's wing. This results in effectively a mixed state made up of $|\uparrow\rangle_{z}$ and $|\downarrow\rangle_{z}$ spin components (with equal weighting) in Bob's wing. Let us first focus on the particles with the $|\uparrow\rangle_{z}$ spin component occurring in Bob's wing. For a particle in the spin state $|\uparrow\rangle_{z}$  passing through a non-ideal SG setup involving an inhomogeneous magnetic field along the z-axis, the probability of such a particle being post-selected in the upper half of the y-z plane is given by $1-E_{s}$ where, as seen from Eq. (\ref{t3b}), the quantity $E_{s}$ denotes the time-saturated probability of finding $|\uparrow\rangle_{z}$ particles in the lower half of the y-z plane ($E_{s}\neq0$ due to non-idealness of the SG 
 setup). The post-selected normalised spin state in this case is $|\uparrow\rangle_{z}$.

Then, for such post-selected particles, using the expression  for the spin state $|\uparrow\rangle_{z}$ in terms of the eigenstates of $\sigma_{\theta}$ with eigenvalues +1 and -1 respectively, the probability of obtaining a particular outcome, say +1, of measuring $\sigma_{\theta}$ is given by

\begin{equation}
\label{t8}
p_{B}^{+}=(1/2)(1+cos\theta)
\end{equation}

Now, remembering that due to measurements of the z-component of spin in Alice's wing, particles with the $|\uparrow\rangle_{z}$ spin component are produced in Bob's wing with the probability (1/2), the probability $P_{B}^{+}$ that such particles passing through the non-ideal SG setup get post-selected in the upper half of the y-z plane and yield the outcome +1 for the measurement of $\sigma_{\theta}$ is given by $P_{B}^{+}=(1/2)(1-E_{s})p_{B}^{+}$ where $p_{B}^{+}$ is given by Eq. (\ref{t8}), whence

\begin{equation}
\label{t9}
P_{B}^{+}=(1/4)(1+cos\theta)(1-E_{s})
\end{equation}

Next, there is another set of particles having the $|\downarrow\rangle_{z}$ spin component in Bob's wing occurring with the probability 1/2 due to measurements of the z-component of spin in Alice's wing. The probability $P_{B}^{-}$ of such particles getting post-selected in the upper half of the y-z plane and yielding the outcome +1 for the measurement of $\sigma_{\theta}$ is as follows (obtained in a way similar to the derivation of the expression (\ref{t9}) for the quantity $P_{B}^{+}$)

\begin{equation}
\label{t10}
P_{B}^{-} = (1/4)(1-cos\theta)E_{s}
\end{equation}

\noindent where, as seen from Eq. (\ref{t3a}), the quantity $E_{s}$ denotes the time-saturated probability of finding $|\downarrow\rangle_{z}$ particles in the upper half of the y-z plane.

Thus, in the case B, in Bob's wing, the total probability $P_{B}^{z}$ of obtaining the outcome +1 for the measurement of $\sigma_{\theta}$ on particles that pass through the non-ideal SG setup and are selected in the upper half of the y-z plane is given by $P_{B}^{z}= P_{B}^{+} + P_{B}^{-}$ where $P_{B}^{+}$ and $P_{B}^{-}$ are given by Eqs. (\ref{t9}) and (\ref{t10}) respectively, whence

\begin{equation}
\label{t11}
P_{B}^{z} = (1/4) \left[1 + (1 - 2E_{s}) cos\theta\right]
\end{equation}

\noindent where the superscript z is used to denote that the quantity $P_{B}^{z}$ is measured in Bob's wing corresponding to the total set of measurements of the z-component of spin in Alice's wing.\\

Now, in this example, the condition ruling out any possibility of
signalling from Alice to Bob that could have occurred by comparing
the cases A and B is given by  $P_{A}^{x} = P_{B}^{z}$,
with this equality holding good for the measurement of any
arbitrary spin component $\sigma_{\theta}$ on the particles in
Bob's wing that are post-selected following a non-ideal SG setup. Using Eqs. (\ref{t7})
and (\ref{t11}), the \textit{no-signalling condition} $P_{A}^{x} = P_{B}^{z}$ reduces to

\begin{eqnarray}\nonumber
\label{t12}
(1/4)\left[1+(1-2E_{s})cos\theta + \sqrt{E_{s}(1-E_{s})}sin\theta\right.\\  
\left.(cos\phi_{+x}+cos\phi_{-x})\right] =(1/4)\left[1+(1-2E_{s})cos\theta\right]
\end{eqnarray}

For $\theta = 0$, i.e., if for the post-selected particles, the spin component is measured along the z-axis (which is the direction of the inhomogeneous magnetic field in the non-ideal SG setup), Eq. (\ref{t12}) is automatically satisfied. For any value of $\theta\neq0$, if the equality given by Eq. (\ref{t12}) is to hold good, the condition $cos\phi_{+x} + cos\phi_{-x} = 0$ needs to be satisfied, which leads to the following relation

\begin{equation}
\label{t13}
\phi_{+x}\pm \phi_{-x} = \pi
\end{equation}

In the above derivation of Eq.
(\ref{t13}), the case A pertains to measurements of the
x-component of spin in Alice's wing,while in the case B, a key feature is that measurements of the spin component in Alice's wing are considered
along a direction (viz. the z-axis) which is the same as that of
the inhomogeneous magnetic field in the non-ideal SG setup in
Bob's wing.

Let us now discuss what happens if in the case A, in
Alice's wing, an arbitrary spin component $\sigma_{\omega}$ is
measured where the angle $\omega$ specifies a direction with
respect to the z-axis in the x-z plane (with the proviso
$\omega\neq 0,\pi$). Then, if one
follows the line of calculation similar to that in the earlier case
A, the total probability $P_{A}^{\omega}=P_{A}^{+}(\omega) + P_{A}^{-}(\omega)$ of obtaining the outcome +1 for the measurement of $\sigma_{\theta}$ on particles in Bob's wing that pass through a non-ideal SG setup and are selected in the upper half of the y-z plane is given by

\begin{eqnarray}\nonumber
\label{t14}
P_{A}^{\omega} = (1/4)\left[1 + (1-2E_{s}) cos\theta + \sqrt{E_{s}(1-E_{s})}\right.\\
\left.sin\omega sin\theta (cos\phi_{+\omega} + cos\phi_{-\omega})\right]
\end{eqnarray}

\noindent where the superscript $\omega$ is used to denote that here the quantity $P_{A}^{\omega}$ measured in Bob's wing corresponds to measurements of the spin component $\sigma_{\omega}$ in Alice's wing. Note that Eq. (\ref{t14}) reduces to Eq. (\ref{t7}) if the measurement of the x-component of spin is perfomed in Alice's wing; i.e., when $\omega = \pi/2$.

Using Eqs. (\ref{t11}) and (\ref{t14}), the \textit{no-signalling
condition} (NSC) $P_{A}^{\omega} = P_{B}^{z}$, in this general case, becomes the following equality 

\begin{eqnarray}\nonumber
\label{t15}
(1/4) \left[1+(1-2E_{s})cos\theta + \sqrt{E_{s}(1 - E_{s})}sin\omega sin\theta\right.\\
\left. (cos\phi_{+\omega} + cos\phi_{-\omega})\right] = (1/4)\left[1+(1-2E_{s})cos\theta\right]
\end{eqnarray}

For any value of $\theta \neq 0$, if Eq. (\ref{t15}) is to be valid, it is required that $cos\phi_{+\omega} + cos\phi_{-\omega} = 0$ which, in turn, implies the following relation

\begin{equation}
\label{t16}
\phi_{+\omega} \pm \phi_{-\omega} = \pi
\end{equation}

As a consequence of NSC, Eq. (\ref{t16}), therefore, provides a relation constraining the relative phase occurring in the spin state of the particles in Bob's wing that are post-selected following a non-ideal SG setup, confined to the upper half of the y-z plane. For $E_{s} = 0$, i.e., when the SG setup used in Bob's wing is \textit{ideal}, it is seen from Eq. (\ref{t15}) that the NSC condition is automatically satisfied. Thus, \textit{non-idealness} of the SG setup ($E_{s}\neq0$) used in our example is crucial for obtaining the quantitative relation given by Eq. (\ref{t16}). The way such a relation can be subjected to an experimental test will now be discussed as follows.

\section{empirical testability of the constraint relation Eq. (\ref{t16})} 

We begin by noting that a possible test of NSC would seem to be by measuring the statistics of the outcomes in one of the two wings of an EPRB setup while changing the measurement setting in the other wing. However, such a test would require to ensure strict spacelike separation between the two relevant measurements in the two wings of the EPRB pairs - a condition which is non-trivial to satisfy because of an ambiguity concerning the stage at which a measurement process can be regarded as completed (i.e., precisely when a measurement outcome can be considered to be registered); this would critically depend upon the details of how a measurement process is modelled. On the other hand, an important point to be stressed is that although the preceding treatment in our paper deriving Eq. (\ref{t13}) or (\ref{t16}) from NSC is in the context of EPRB entangled pairs, once the relation is obtained, its validity can be studied by focusing only on a single beam of spin-1/2 particles (say, neutral atoms or neutrons) passed through a non-ideal SG setup. Thus, in practice, such a test wouldn't require EPRB pairs, thereby circumventing the need to satisfy the delicate condition of spacelike separation.

We first consider a beam of spin-1/2 particles with their spins oriented along a direction making an angle $\omega$ with the z-axis in the x-z plane (here $\omega\neq 0, \pi$). Let this beam having the spin state $|\nearrow\rangle_{\omega}$ be passed through a non-ideal SG setup in which the inhomogeneous magnetic field is along the z-axis. Subsequently, attention is focused on the spin state of a set of particles confined to the upper half of the y-z plane that are selected for further measurement by blocking/detecting particles in the other half of the y-z plane. Such a spin state is of the form given by Eq. (\ref{t2}) where the relative phase is $\phi_{+\omega}$ for the initial beam of spin-polarized particles with the spin state $|\nearrow\rangle_{\omega}$. Similarly, if a beam of oppositely spin-polarized particles with the spin state $|\swarrow\rangle_{\omega}$ is passed through the non-ideal SG setup in question, the spin state of the post-selected particles given by Eq. (\ref{t2}) would involve the relative phase $\phi_{-\omega}$.

Now, note that the relative phase occurring in the spin state of the particles selected in, say, the upper half of the y-z plane cannot be fixed unless the effect of detecting particles in the other half is taken into account within a fully unitary treatment. Whatever be the details of such a treatment, the upshot of our preceding analysis is that the sum of the phase factors $\phi_{+\omega}$ and $\phi_{-\omega}$ should turn out to be $\pi$ as constrained by Eq. (\ref{t16}) which is a consequence of the no-signalling condition. Therefore, an empirical test of the constraint relation given by Eq. (\ref{t16}) would provide a means for precision testing of NSC. 

Next, to see explicitly how the experimental determination of $\phi_{+\omega}(\phi_{-\omega})$ required for testing Eq. (\ref{t16}) can be realized with respect to the post-selected spin state of the form given by Eq. (\ref{t2}),  we write this state as follows

\begin{equation}
\label{t17}
|\nu\rangle = \sqrt{1-E_{s}}|\uparrow\rangle_{z} + exp(i\phi_{\pm\omega})\sqrt{E_{s}}|\downarrow\rangle_{z}
\end{equation}

Given a representative set of post-selected particles corresponding to the above state $|\nu\rangle$, if one considers the measurement of the spin variable $\sigma_{z}$, it is evident from Eq. (\ref{t17}) that the probability of obtaining the outcome $\pm1$ will yield the value of the parameter $E_{s}$. Then, if one takes another representative set of such particles and considers the measurement of any spin component, say $\sigma_{x}$, non-commuting with $\sigma_{z}$, the probability of obtaining the outcome, say +1, evaluated using Eq. (\ref{t17}) written in terms of the eigenstates of $\sigma_{x}$, will be given by $(1/2) + \sqrt{E_{s}(1-E_{s})} cos\phi_{\pm\omega}$. This measured probability, therefore, enables to fix the phase factor $\phi_{\pm\omega}$, since the parameter $E_{s}$ is known from the measurement of $\sigma_{z}$. Thus, in this way, by determining $\phi_{\pm\omega}$, the NSC relation given by Eq. (\ref{t16}) can be subjected to an experimental verification. 

Note that the accuracy of the above experimental determination of $\phi_{\pm\omega}$ depends on the accuracy to which the \textit{idealness} of the SG setup is ensured in measuring the relevant spin variables pertaining to the post-selected spin state, while the parameters of the \textit{non-ideal} SG setup used before post-selection need to be chosen such that the quantity $E_{s}$ has an appreciable non-zero value. As mentioned above, the measured probability from which $cos\phi_{\pm\omega}$ can be calculated involves the factor $\sqrt{E_{s}(1-E_{s})}$ which will determine the overall precision to which NSC can be tested using the method formulated in this paper and an empirical upper bound on possible violation of NSC can be provided.

We would also like to point out that in our example, the relevant parameters of the SG setup and the region of the position space over which one introduces the projection/post-selection determine the probability amplitudes in the superposition given in Eq. (\ref{t2}) (or, Eq. (\ref{t17})) for the post-selected spin state, through the parameter $E_{s}$ fixed by time-saturated values of Eqs. (\ref{t3a}) and (\ref{t3b}). A key point of our treatment is that whatever be the value of the parameter $E_{s}$, the constraint relation Eq. (\ref{t16}) must hold good.

\section{summary and Conclusion}
We would like to stress that our derivation of the relation (\ref{t13}) or (\ref{t16}) is based on NSC and the standard rules of quantum mechanics that include, in particular, taking the overall dynamics of the non-ideal Stern-Gerlach device to be linear and the probabilities for measurement outcomes at any given time given by the standard Born rule. If the relation (\ref{t13}) or (\ref{t16}) is found to be empirically violated, it would then seem to imply violation of NSC for the following reason. As mentioned earlier in Section I by citing Ref. [10], Born's rule has recently been verified by a rigorous precision test. On the other hand, the linearity condition of quantum dynamics and NSC are interlinked \cite{gisin} and, in particular, the thorough analysis by Simon et al.\cite{simon} brings out the point that once NSC and the experimentally well-verified standard Born rule are taken to be valid, quantum dynamics is rather rigidly constrained to be linear. Thus, if the relation (\ref{t13}) or (\ref{t16}) turns out to be empirically  invalid, it would seem plausible to infer that the violation of NSC arises from a departure from linearity in the way a non-ideal Stern-Gerlach setup acts in conjunction with the type of post-selection process that is used in our example. The tenability of such a possibility would then call for further investigation; for example, the general framework for accommodating non-linearity in quantum dynamics discussed by Weinberg \cite{weinberg} may be invoked in the context of our setup in order to explain any observed violation of the relation (\ref{t13}) or (\ref{t16}). Here we note that in Weinberg's paper, an indication \cite{weinberg2} has been given of applying his general framework by analysing an ideal Stern-Gerlach setup, while in our example, a non-ideal Stern-Gerlach device is considered, along with a suitable post-selection of particles emerging in one of the two halves.\\

As regards the distinctiveness of the setup used in the present
paper, we may stress that the post-selection procedure
used here following a non-ideal SG setup is \textit{different} from the
scheme of weak-measurement related studies \cite{aharonov} where the post-selection follows an ideal SG setup that is preceded by a non-ideal SG
with a very weak magnetic field. In our example, the central
feature is that for an incoming beam of spin-1/2 particles that are spin-polarized along any direction (other than the z-axis, the direction of the magnetic field in the non-ideal SG setup considered here), the post-selected spin state of the emerging particles in any one of the two halves is a superposition of
$|\uparrow\rangle_{z}$ and $|\downarrow\rangle_{z}$ states with a relative phase for which a constraint relation is obtained. 

It has  already been mentioned that a complete unitary treatment is required for evaluating the above mentioned relative phase, for which one would need to suitably incorporate the effect on the post-selected spin state in one of the halves, arising from the blocking/detecting of particles in the other half that emerge from the non-ideal SG setup. The extent to which the details of the modelling of such a post-selection process can affect the evaluation of the relative phase occurring in the post-selected spin state should be instructive to probe. Importantly, this type of study needs to be compatible with the constraint relation given by Eq. (\ref{t13}) or (\ref{t16}) whose validity should be independent of the specifics of the modelling of the post-selection procedure used in our example. Here it may be noted that, given the theory of approximate or generalized
measurement \cite{busch} that has been developed to a considerable extent, its
possible implication pertaining to the type of post-selection considered here could be worth investigating. Finally, it is hoped that the predictive power of NSC in a testable
situation as illustrated by the example treated here using a non-ideal measurement setup may motivate the formulation of other such examples that could be helpful for a deeper understanding of the role of NSC in the context of non-ideal measurement situations.

\section{Acknowledgements} 
DH is thankful to John Corbett and Paul Davies for interactions that motivated this work. We thank Paul Busch, Harvey Brown, Alex Matzkin and Tony Sudbery for useful comments concerning the preliminary draft of this paper. We also thank the referees for helpful suggestions. Support from the DST Project No. SR/S2/PU-16/2007 is acknowledged. DH also thanks Centre for Science, Kolkata for supporting his research.


\end{document}